\documentstyle[eqsecnum,aps]{revtex}
\begin{document}
%%\draft
\title{
Pathlength distributions of atmospheric neutrinos
}
\author{ T.K. Gaisser \& Todor Stanev}
\address{
 Bartol Research Institute, University of Delaware, Newark, DE 19716
}
%\date{\today}
\maketitle
\begin{abstract}
We present the distribution of the production heights of
atmsopheric neutrinos as a function of zenith angle and 
neutrino energy.  The distributions can be used as the
input for evaluation of neutrino propagation under various
hypotheses for neutrino flavor oscillations.
\end{abstract}

\section{Introduction}

 Initial results from SuperKamiokande \cite{SuperK} appear
 to confirm indications from IMB,~\cite{IMB} Kamiokande~\cite{Kam}
 and Soudan~\cite{Soud} of an excess of $\nu_e$ relative to $\nu_\mu$
 in the atmospheric neutrinos. One possible interpretation is
 that neutrino flavor oscillations play a role. In a two--flavor
 mixing scheme, for example, the probability that a neutrino
 of flavor $i$ and energy $E_i$ retains its identity after
 propagating a distance $L$ in vacuum is \cite{BoehmV}
\begin{equation}
 P_{ii}\;=\;1\,-
 \,\sin^22\theta\,\sin^2\left[1.27\,\Delta m^2(eV^2)\times L(km)\over
                                   E_i(GeV)\right],
\end{equation}
 where $\delta m^2$ is the difference in mass squared of the
 two neutrino mass eigenstates and $\theta$ is the mixing angle.
 Therefore, to evaluate the manifestation of the mixing
 in a detector that measures to some degree the direction
 and energy of neutrino-induced events, one needs to know
 the distributions of production heights of the neutrinos
 as a function of energy and zenith angle. More complicated
 mixing schemes~\cite{Foglietal} and effects of propagation
 in matter~\cite{Parkeetal} still require this basic information
 about the points of origin of the neutrinos.

 Information about origin of the neutrinos is implicit in any
 calculation of neutrino fluxes.  Here we extract the relevant
 information from the simulation of Ref. \cite{Agrawal},
 which has been compared to several other calculations in
 Ref.~\cite{GHKLMNS}.

 The paper is organized in three sections.  First we review the
 simulation we are using to calculate production of neutrinos
 in the atmosphere.  Next we present the basic results of the
 calculation. We discuss simple analytic approximations which
 offer insight into the systematics of the results and compare
 them to simulation results for zenith angles from the vertical to
 horizontal. Finally, we provide some parametrizations, based
 on the analytic approximations, that may be useful for practical
 application of the results.

\section{Simulation}

  The simulation was performed in the spirit of earlier
 calculations of the atmospheric neutrino flux~\cite{Agrawal,BGS}.
 The simulation code is one dimensional.  In this approximation,
 all secondaries are assumed to move in the direction of the
 primary particles (except for a small fraction of low energy
 secondaries  with angles larger than $90^\circ$ to the beam,
 which are discarded).  The validity of this approximation has
 been checked in Refs. \cite{Lee,conference}.
 
  The primary cosmic ray flux and its composition is the
 parametrization used previously in the calculation of Agrawal
 {\em et al.}~\cite{Agrawal} which in the  multi-GeV range falls in
 between the measurements of Refs.~\cite{Webber,Seo}.   Incident
 cosmic-ray nuclei are treated in the superposition approximation \cite{EGLS},
 with cascades generated separately for protons and neutrons in
 order to insure the correct ratios of neutrinos and antineutrinos.
 The fraction of neutrons is derived from the 
 fractions of nuclei heavier than hydrogen in the primary flux. 

 We consider three ranges of neutrino energies that correspond
 approximately to the three major types of experimental events
 in a detector the size of SuperKamiokande:
 contained events; partially contained neutrino interactions
 and stopping neutrino induced muons; and througoing muons.
 The energy ranges are presented in two different ways:
\begin{itemize}
\item $0.3<E_\nu<2$~GeV; $2<E_\nu<20$~GeV; $E_\nu>20$~GeV and
\item $E>1,\;10$~and~$100$~GeV.
\end{itemize}
The integral form is more closely related to simple analytic
approximations that we use as the basis of parametrizations of
the results of Monte Carlo simulations.

 Our results are obtained with the geomagnetic cutoffs for
 Kamioka and for the epoch of solar minimum, which is applicable
 to measurements performed currently ($\sim1994-99$).
 Because of the high geomagnetic cutoffs at Kamioka, it is not
 necessary to account precisely for the phase of the solar cycle.
To illustrate the potential influence of geomagnetic effects
at other locations we also tabulate some results for the much higher
 geomagnetic latitude of the SNO experiment.

  We have not included  prompt neutrino production through
 charm decay because it is totally negligible in the considered
 energy ranges~\cite{ThunIng,Volkova}. All neutrinos are
 generated either in pion and kaon decays or in muon decays.
 The production heights are stored separately for neutrinos
 from $\pi/K$ and from muon decays. The muon decay procedure
 accounts for the muon enegy loss during propagation in the atmosphere.
 Technically the muon lifetime is sampled in the muon rest
 frame and then the muon is propagated in the atmosphere with time
 dilation proportional to its decreasing energy. Thus mouns decay
 on the average sooner that they would have if one (incorrectly)
 sampled from a decay distribution using their Lorentz factor
 at production.

\section{Results}

  Before presenting the results for neutrinos, we show
 a comparison between measurements of GeV muons 
 at different altitudes in the atmosphere and our calculation made
 with the same Monte Carlo code \cite{Circella}. This type of
 balloon measurement provides the most direct
test of the validity of the cascade model and of
 the treatment of the muon propagation in the atmosphere because
the muons and neutrinos have a common origin. The
comparison shown in Fig. 1 is with data of the MASS experiment \cite{MASS}
as discussed in Ref. \cite{Circella}.

  Fig.~2 shows the height of production of neutrinos of energy
 above 1 GeV for cos$(\theta)$ = 0.75. The graph gives $dN_\nu/dh$
 (cm$^{-2}$s$^{-1}$sr$^{-1}$km$^{-1}$), where $h$ is the slant distance
 from the neutrino production point to sea level.  
 Contributions from muon decay and from $\pi/K$ decay are shown separately for
 $\nu_e + \bar{\nu}_e$ and for $\nu_\mu + \bar{\nu}_\mu$.
 The overall flux of
 $\nu_e+\bar{\nu}_e$ from $\pi/K$ decay is much lower because it
 reflects primarily the contribution of $K^0_L$ decays, which is very low
 in this energy range. The curves for electron and for muon
 neutrinos from muon decay are nearly equal. They extend to lower
 altitudes with a slope that depends on the average energy of
 the parent muons. For higher energy neutrinos this slope is
 signicantly flatter as a consequence of the higher parent muon
 energy and correspondingly longer muon decay length. For $E_\nu >$
 20 GeV, most parent muons reach the ground (except in nearly
 horizontal direction) and stop before decaying.  As a consequence,
the height distributions for neutrinos from muon decay deep in the
atmosphere are nearly flat.

\subsection{Height distribution for neutrinos from $\pi/K$ decay}

\subsubsection{Analytic approximation}

 It is instructive to look at a simple approximation
 for the height of production of neutrinos from decay of pions.
 In the approximation of an exponential atmosphere with scale height
 $h_0$ and the approximation of Feynman scaling for the production
 cross sections of pions in interactions of hadrons with nuclei
 of the atmosphere, a straightforward solution of the equations
 for propagation of hadrons through the atmosphere \cite{book} 
 gives \cite{Lipari}
\begin{equation}
{dF(>E_\nu)\over dX}\;=\;(1-r_\pi)^{-\gamma}\,{Z_{N\pi}\over\lambda_N}\,
e^{-X/\Lambda_N}\,{K\over\gamma(\gamma+1)}\,E_\nu^{-\gamma}\;
\equiv\;A \times E_\nu^{-\gamma}
\label{production}
\end{equation}
 for the integral flux of neutrinos in the energy range
 $E_\nu\ll \epsilon_\pi$ where reinteraction of pions in the
 atmosphere can be neglected.  There is a similar expression
 for neutrinos from decay of kaons proportional to
 $B_K\times Z_{NK}$. The meaning and approximate
 values of the quantities in these equations are given in
 Table~\ref{tab1}.

\begin{table}
\caption{Values of the parameters used in Eq.~\ref{production} that
 correspond to a power law primary cosmic ray spectrum and to an
 exponential atmosphere. $\gamma$ and $K$ are the spectral index
 and the coefficient of the differential cosmic ray energy spectrum,
 $dN/dE = K E^{-(\gamma+1)}$.  $r_\pi \; (r_K)$ is $(m_\mu/m_\pi)^2\;
 ((m_\mu/m_K)^2)$. $Z_{N\pi}\;(Z_{NK})$ is the spectrum weighted
 moment for pion (kaon) production by nucleons
 ($Z_{N\pi}\; =\; \int\,dx\,x^\gamma\,dN/dx$). $\Lambda_N$ and $\lambda_N$
 are the attenuation and interaction lengths for nucleons. $B_K$ is the
 branching ratio for $K \longrightarrow \mu$ decay. $X_0$ and $h_0$
 are the total vertical thickness (in g/cm$^2$) and the scaleheight
 for an exponential atmosphere in km.}
\begin{tabular}{cccccccccc}
 $\gamma$ & $K$ & $\lambda_N$ & $\Lambda_N$ & $Z_{N\pi}$ &$B_K \times Z_{NK}$ &
 $r_\pi$  & $r_K$ & $X_0$ & $h_0$ \\
  & cm$^{-2}$s$^{-1}$sr$^{-1}$(GeV)$^\gamma$ & g/cm$^2$ & g/cm$^2$ & & &
          &       & g/cm$^2$ & km \\ \tableline
 & & & & & & & & & \\
  1.70 & 1.8 & 86 & 120 & 0.08 &0.0075  &   0.5731 & 0.0458 & 1030 & 6.4 \\
 & & & & & & & & & \\ \tableline
\end{tabular}
\label{tab1}
\end{table}

 In Eq.~\ref{production} $X$ is the slant depth in the atmosphere
 at which the pion is produced and decays.  We now convert this
 into distance $\ell$ from the detector in the approximation of
 an exponential atmosphere in which 
\begin{equation}
X\;=\;{X_0\over \cos\theta}\,\exp\left[-{\ell\cos\theta \over h_0}\right].
\label{atmos}
\end{equation}
 For $\theta<70^\circ$, curvature of the earth can be neglected
 and $\cos\theta$ in Eq.~\ref{atmos} is cosine of the zenith
 angle to a good approximation.  The effective values of $\cos\theta$
 for larger angles are given below.

 The corresponding approximate expression for the distribution
 of production distances is
\begin{equation}
{dF(>E_\nu)\over d\ell}\;=\;{A X_0\over h_0}\, E_\nu^{-\gamma}
   \exp\left[-{X\over \Lambda_N}\right] 
       \times\exp\left[-{\ell\cos\theta\over h_0}\right],
\label{nupi}
\end{equation}
 where $X$ is to be evaluated as a function of $\ell$ from
 Eq.~\ref{atmos}. Assuming a primary cosmic ray nucleon flux
with the normalization given in Table \ref{tab1} (and including
the small contribution from decay of kaons) the normalization
 factor is $A X_0/h_0\; \simeq$ 0.020.

 Taking parameters from Table~\ref{tab1} gives the most probable
 distance of production as 
\begin{equation}
\label{mode}
\ell_{max}\;\approx\;{h_0\over\cos\theta}\,\ln{X_0\over\Lambda_N\cos\theta},
\end{equation}
 which is $\approx 15$~km for vertical neutrinos from decay of pions.

\subsubsection{Monte Carlo results}

  Fig.~3 shows the distance distribution for neutrinos from $\pi/K$ decay
 ($ E_\nu >$ 1 GeV) for cos$(\theta)$ = 1.00, 0.75, 0.50, 0.25,
 0.15 and 0.05. Here $\theta$ is the zenith angle of the
neutrino trajectory at the surface of the Earth.  
 For large zenith angles, the curvature of the Earth is significant,
and it is necessary to use effective values of 
$\cos_{eff}(\theta)$ that represent the convolution
of the locations of neutrino production with
the local zenith angle as it decreases moving upward along the trajectory.
We treat $\cos_{eff}(\theta)$ as a free parameter
in fitting Eqs. \ref{nupi} and \ref{numu} to the Monte Carlo results.
The values are included in Table \ref{tab2}.

\begin{table}
\caption{ Comparison of analytic and Monte Carlo values of the effective
 value of $\cos \theta$ . Column 1 shows the cosine of the zenith angle
 $\theta$. Column 2 shows the most probable production height
 for neutrinos from $\pi/K$ decay from the Monte Carlo calculation. 
 Column 3 gives the most probable height of production from Eq.~\ref{mode}
 with $cos_{eff} \theta $ from column 4.
 Columns 4 \& 5 give the $\cos_{eff} \theta$ values that fit best the
 calculated height of production distributions for neutrinos from $\pi/K$
 and muon decay with $h_0$ = 6.50 km. Column 6 gives the normalization
 coefficient $C_\mu$ needed to fit the distribution for neutrinos from
 muon decay.}
\begin{tabular}{lccllc}
 $\cos \theta$ & $\ell_{max}(MC)$ &  $\ell_{max}$(Eq.~\ref{mode}) &
 $\cos_{eff}^{K/\pi} \theta$ & $ \cos_{eff}^\mu \theta$ & $C_\mu$ \\
               & (km) & (km) & & & \\
 \tableline
 & & & & & \\
 1.00 & 13.8 & 14.0 & 1.00 & 1.00   & 0.69 \\
 0.75 & 21.6 & 21.2 & 0.75 & 0.75   & 0.71 \\
 0.50 & 38.4 & 37.0 & 0.50 & 0.50   & 0.77 \\
 0.25 & 88.4 & 87.5 & 0.26 & 0.26   & 0.83 \\
 0.15 & 155. & 157. & 0.164 & 0.168 & 1.00 \\
 0.05 & 382. & 358. & 0.084 & 0.087 & 1.86 \\
 & & & & & \\
\tableline
\end{tabular}
\label{tab2} 
\end{table}  
  
  Up to $\cos\theta$ = 0.25 the agreement between the Monte Carlo
 calculation and the analytic estimate is quite good.
 Note that  for nearly horizontal neutrinos the height distribution
 from the Monte Carlo calculation is artificially narrow  and irregular.
 The atmospheric model used does not treat exactly the atmospheric
 densities at vertical depths of less than few g/cm$^2$.  This intruduces
 a sharp cutoff in the height distribution for strongly inclined showers
 and also decreases the width of the height distribution.

  The height distribution of $\nu_e$ from $\pi/K$ decay has a similar shape
 with much lower normalization, because only $K^0_L$ have decay
 mode with $\nu_e$'s ($K^0_{e3})$.

\subsection{Height distribution for neutrinos from muon decay}

\subsubsection{Analytic approximation}

To estimate the height of production for neutrinos from decay
of muons is more complicated because of the competition between
decay and energy loss for muons in the multi-GeV energy range.
One starts from the distribution of production points for muons,
which is similar to Eq. \ref{production} with different coefficients.
The resulting approximate expression  \cite{Lipari}
for the distribution of production distances (differential in
the energy of the parent muons as well as the slant height of production) is
\begin{equation}
\label{numu}
{dN_\nu\over dE_\mu\,d\ell}\;=\;K B {\mu c^2\over E_\mu c\tau}\,\int_0^X\,
{dY\over \lambda_N}{e^{-Y/\Lambda_N}\left[{X \over Y}{E_\mu+\alpha(X-Y)\over
E_\mu}\right]^{-p}\over[E_\mu+\alpha(X-Y)]^{\gamma+1}},
\end{equation}
where $\tau$ is the muon lifetime,
$$
p\;=\;{h_0\over c\tau\cos\theta}{\mu c^2\over E_\mu+\alpha X}
$$
and
$$
B\; = \; {1\over\gamma+1}\,\left[{1-r_\pi^{(\gamma+1)}\over 1-r_\pi}\,Z_{N\pi}\;
+\;B_{K}\,{1-r_k^{(\gamma+1)}\over 1-r_K}\,Z_{NK}\right].
$$
At high altitude muon energy loss ($\alpha(X-Y)$) can
be neglected and the expression \ref{numu} is proportional to
slant depth $X$ given by Eq. \ref{atmos}.  This expression
gives a good account of the high-altitude exponential falloff
of the neutrinos from muon decay. 

An approximation that is adequate for fitting the distribution
for all distances (integrated over neutrino energy) is
\begin{equation}
\label{numuapprox}
{dN_\nu(>E_\nu)\over d\ell}\,\approx\,
{C_\mu\,K\,B\over (\gamma+1) (2 E_\nu)^{(\gamma+1)}}\,
{\mu c^2\over c\tau}\,{X\over\lambda_N}\,
\int_0^1 dz\,z^p\,\exp{(-{X\over\Lambda_N}z)}
\left[1\,+\,{\alpha X\over 2\,E_\nu}(1-z)\right]^{-(p+\gamma+1)} ,
\end{equation}
where $C_\mu$ is an overall normalization factor used to fit
the Monte Carlo results (see Table~\ref{tab2}).

\subsubsection{Monte Carlo results}

 Fig.~4 shows the height of production distributions for muon
 neutrinos of energy above 1 GeV from muon decay. The lines
 are calculated  according to Eq.~\ref{numuapprox} with values of
$\cos_{eff}\theta$ as given in Table~\ref{tab2}.  To obtain the
fits shown in Fig.~4 the approximations of Eq. \ref{numuapprox}
have also been renormalized as indicated in Table~\ref{tab2}.

 At high altitude the height of production distributions have the same
 shape as the ones from neutrinos from $\pi/K$ decay, shifted to lower
 altitudes by one muon decay length (6.24 km for 1 GeV muons). 
 At lower altitude the shapes are quite different. The production
 height for neutrinos from muon decay extend to much lower altitude
 because of the slow attenuation of the parent muon flux, an effect which
 becomes more pronounced as the energy increases.

  It is interesting to observe that at high zenith angles the yield
 of neutrinos from (daughter) muon decay exceeds the yield of neutrinos
 from the decay of the parent pions and kaons. 
 The reason is that muon neutrinos
 from muon decay in flight have a spectrum extending almost to
 $x = 1$ (where $x = E_{\nu}/E\pi$), while the neutrinos from $\pi$
 decay can only reach $E_\nu^{max} = E_\pi \times (1 - r_\pi)$
 = $0.428\,E_\pi$. The corresponding $Z$--factors $Z_{\pi\mu \nu_\mu}$
 and $Z_{\pi\nu_\mu}$ are 0.133 and 0.087 respectively, including
the effect of  muon polarization in pion decay. The result is that
for large zenith angles, when almost all muons decay the
$\nu_\mu$ yield from muon decay becomes slightly larger than
that from $\pi/K$ decay ($\sim 9/7$ for $\cos\theta$ = 0.05).

  Fig.~5 compares the distributions of distance to production
for $\nu_\mu$
 from muon decay with $E_\nu$ above 1, 10, and 100 GeV.  At high 
 neutrino energy the muon decay length becomes comperable or larger
 than  the total dimension of the atmosphere. The neutrino height of
 production then becomes constant deep in the atmosphere.

  The height distribution for $\nu_e$ from muon decay is analogous
 to that of $\nu_\mu$. The only difference is the slightly lower
 normalization, which reflects the ratio $Z_{\mu \nu_\mu}/Z_{\mu \nu_e}$
 = 0.133/0.129 = 1.03 (including muon polarization).

\subsection{Height distribution in three energy bins}

  In Table~\ref{tab3} we show the average height of production and
 the contributions of $\pi/K$ and muon decays for neutrinos in the
 three energy bins ($0.3 <\,E_\nu\,< 2$~GeV; $2 <\,E_\nu\,< 20$~GeV;
 $E_\mu\,20$~GeV) which roughly correspond to contained neutrino
 events, semicontained events and stopping neutrino induced muons and
 throughgoing neutrino induced muons. For each angle and neutrino flavor
 Table~\ref{tab3} first gives the average height of production
 (slant depth) in km and the width of the height of production
 distribution. Then it gives the contribution (in \%) of $\pi/K$ decay
 $f_m$ and the corresponding $\langle h_m \rangle$ and $\sigma h_m$,
 then the same quantities ($f_\mu,\; \langle h_\mu \rangle,\; \sigma h_\mu$
 for neutrinos from muon decay.

  The calculation was done with the geomagnetic cutoffs of Kamioka, 
 except for the three lines ($\cos \theta$ = 1.00, 0.75 and 0.50, for
 the lowest energy bin) that are also calculated for
 the high geomagnetic latitude of SNO~\cite{SNO}. The numbers for 
 high geomagnetic latitude are slightly higher (2 -- 10 \%) for
 both neutrino sources because of the contribution of low energy
 protons. This difference becomes negligible at higher angles. 

\begin{table}
\caption{ Production height (slant distance, km) of neutrinos for six
 values of $\cos{\theta}$ and three neutrino energy ranges. The calculation
 is for the geomagnetic location of Kamioka with three lines for the lowest
 energy range calculated for Sudbury, Canada.}
\begin{small}
\begin{center}
\begin{tabular}{|r||rr|rrr|rrr||rr|rrr|rrr|}\hline
 E, GeV & \multicolumn{8}{ c|}{$\nu_e + \bar{\nu}_e$} &
          \multicolumn{8}{|c|}{$\nu_\mu + \bar{\nu}_\mu$} \\ 
 $\cos{\theta}$ &h&$\sigma_h$&$f_{\pi/K}$&$h_{\pi/K}$&$\sigma h_{\pi/K}$  
                                 & $f_\mu$ & $h_\mu$ & $\sigma h_\mu$ &
                 h&$\sigma_h$&$f_{\pi/K}$&$h_{\pi/K}$&$\sigma h_{\pi/K}$  
                                 & $f_\mu$ & $h_\mu$ & $\sigma h_\mu$ \\
\hline \hline
 0.3 -- 2.  & & & & & & & & & & & & & & & &\\
1.00  & 14.0&  8.7&  1.3& 16.8&  8.3& 98.7& 14.0&  8.7&
    15.9&  8.7& 57.4& 17.4&  8.4& 42.6& 14.1&  8.7\\
0.75  & 21.0& 11.9&  1.2& 25.6& 11.9& 98.8& 21.0& 11.9&
    23.6& 11.8& 54.5& 25.6& 11.4& 45.5& 21.1& 11.9\\
0.50  & 37.7& 18.4&  1.0& 44.0& 17.2& 99.0& 37.6& 18.3&
    41.0& 18.1& 51.3& 44.0& 17.2& 48.7& 37.8& 18.3\\
0.25  & 91.1& 32.0&  0.9&100.5& 29.9& 99.1& 90.9& 32.1&
    95.6& 31.4& 48.7&100.2& 30.2& 51.3& 91.3& 32.0\\
0.15  &154.8& 38.1&  0.9&167.6& 35.1& 99.1&154.6& 38.2&
   160.0& 37.3& 47.9&165.6& 35.3& 52.1&155.0& 38.0\\
0.05  &363.8& 56.5&  0.9&378.0& 52.8& 99.1&363.7& 56.5&
   369.8& 55.0& 47.4&376.1& 53.1& 52.6&364.1& 56.3\\
 SNO  & & & & & & & & & & & & & & & &\\
1.00  & 15.4&  8.9&  0.8& 17.2&  8.6& 99.2& 15.4&  8.9&
    16.9&  8.8& 53.8& 18.0&  8.5& 46.2& 15.5&  8.9\\
0.75  & 23.2& 12.1&  0.7& 25.6& 11.3& 99.3& 23.2& 12.1&
    25.0& 11.9& 51.1& 26.5& 11.5& 48.9& 23.3& 12.1\\
0.50  & 40.6& 18.4&  0.6& 44.4& 17.3& 99.4& 40.6& 18.4&
    43.0& 18.1& 48.3& 45.2& 17.4& 51.7& 40.8& 18.3\\
2. -- 20.  & & & & & & & & & & & & & & & &\\
1.00  & 13.4&  9.1&  6.7& 17.9&  8.9& 93.3& 13.1&  9.0&
    16.6&  9.0& 71.6& 18.0&  8.6& 28.4& 13.1&  8.9\\
0.75  & 19.6& 12.4&  5.4& 26.3& 11.6& 94.6& 19.3& 12.3&
    24.1& 12.1& 67.0& 26.4& 11.4& 33.0& 19.4& 12.3\\
0.50  & 34.4& 19.6&  3.9& 44.8& 17.4& 96.1& 34.0& 19.5&
    40.9& 19.1& 60.2& 45.3& 17.5& 39.8& 34.2& 19.3\\
0.25  & 81.9& 35.6&  2.9&102.8& 31.1& 97.1& 81.3& 35.6&
    92.8& 34.6& 52.6&102.9& 30.5& 47.4& 81.7& 35.4\\
0.15  &139.5& 45.2&  2.5&168.6& 34.9& 97.5&138.8& 45.2&
   154.3& 42.8& 49.2&169.1& 35.0& 50.8&139.9& 44.9\\
0.05  &338.9& 73.0&  2.2&380.6& 51.5& 97.8&338.0& 73.1&
   359.0& 67.1& 46.2&381.6& 52.0& 53.8&339.5& 72.4\\
 $>$ 20. & & & & & & & & & & & & & & & &\\
1.00  & 14.0&  9.3& 41.5& 17.7&  8.7& 58.5& 11.4&  8.9&
    17.6&  8.9& 94.2& 17.9&  8.7&  5.8& 11.6&  9.1\\
0.75  & 20.0& 13.1& 33.1& 26.4& 11.7& 66.9& 16.8& 12.6&
    25.8& 12.1& 91.9& 26.6& 11.7&  8.1& 16.6& 12.4\\
0.50  & 31.8& 20.3& 22.3& 44.8& 17.7& 77.7& 28.0& 19.4&
    43.3& 18.9& 87.8& 45.4& 17.8& 12.2& 28.0& 19.4\\
0.25  & 70.3& 38.9&  7.8& 99.1& 27.2& 92.2& 67.2& 38.6&
    94.9& 36.4& 79.6&103.8& 29.9& 20.4& 60.1& 39.5\\
0.15  &110.3& 54.7&  8.8&168.1& 35.0& 91.2&104.7& 53.0&
   151.2& 49.4& 72.5&168.7& 34.6& 27.5&105.2& 52.7\\
0.05  &267.4&105.1&  5.4&382.1& 54.8& 94.6&260.8&103.5&
   335.7& 94.2& 61.7&381.7& 51.1& 38.3&262.3&100.5\\
 \label{tab3}
 \end{tabular}
 \end{center} \end{small} \end{table}

  There are two obvious trends in the numbers in Table~\ref{tab3}.
 The height of production for neutrinos from $\pi/K$ decay grows slightly
 with the neutrino energy because higher energy mesons preferentially
 decay (rather than interact) in the tenuous atmosphere at high altitude.
 Neutrinos from muon decay, on the other hand, are generated at
 lower altitude at high energy because of the increasing muon decay length.
 This second feature is much stronger because of the proportionality
 of decay length and muon energy. 

  The average heights of production also reflect the relative yields of the
 two neutrino sources. For low energy $\nu_e (\bar{\nu}_e)$, for example,
 the contribution of $K^0_{e3}$ is small, so the 
 average height of production is dominated by muon decay.
 At higher energy the relative contribution of $K^0_{e3}$ grows, especially
 at directions close to the vertical, and $\langle h \rangle$ becomes
 intermediate between those of the two processes with correspondingly
 larger width. 

  Generally the contribution from muon decay increases significantly 
 with the zenith angle since even 20 GeV muons easily decay in
 cascades developing in nearly horizontal direction. 

\section{Conclusions}

  We have calculated the distribution of pathlengths of atmospheric
neutrinos for a range of angles and energies relevant for
current searches for neutrino oscillations with atmospheric neutrinos.
Accounting correctly for the pathlenght will be particularly
important for neutrinos near the horizontal direction where the
pathlength through the atmosphere of neutrinos from above the horizon
 is of the same order of magnitude as the pathlength through the Earth
of neutrinos from below the horizon.
  We have also given simple approximations that may be useful
in interpolating the tables and adapting the results for
different energy ranges and directions.

  The influence of the geomagnetic effects on the calculated height
 of neutrino production distributions is not very strong. The difference
 in the average production heights for neutrinos detected at Kamioka
 and SNO is of order several per cent in directions relatively close
 to the vertical. This difference diminishes with angle and becomes
 totally negligible for upward going neutrinos, where the geomagnetic
 cutoffs becomes approximately equal, being averaged over the
 geomagnetic fields of the opposite hemisphere. 

  The agreement of our calculation with the measured muon fluxes above
 1 GeV/c as a function of the atmospheric depth serves as a check
 on the validity  of the results presented above.

 \acknowledgments The authors express their gratitude to 
 W.~Gajewski, J.G.~Learned, H.~Sobel, and Y.~Suzuki for their 
 interest in the height of production problem that inspired us the
 complete this research. This work is supported in part by the U.S.
 Department of Energy under DE-FG02-91ER40626.A007.

\begin{figure}
\caption{Comparison of the calculated flux of negative muons above
 1 GeV/c as a function of the atmospheric depth to the measurements 
 of the MASS experiment.~\protect\cite{MASS,Circella}
\label{fig1}
}
\end{figure}
\begin{figure}
\caption{Height of production distribution for neutrinos of energy
 above 1 GeV at $\cos \theta$ = 0.75. The contributions of $\pi/K$
 and muon decays for the two neutrino flavors are clearly visible. 
 Dots show the height of production for electron neutrinos from
 $\pi/K$ decay, short dashes: $\nu_e + \bar{\nu}_e$ from muon decay.
 The heavy dash line is the sum of the two.  Dash dot: 
 $\nu_\mu +  \bar{\nu}_\mu$ from $\pi/K$ decay, dash--dash:
 $\nu_\mu + \bar{\nu}_\mu$ from muon decay. The heavy solid line is the
 sum of these two.
\label{fig2}
}
\end{figure}
\begin{figure}
\caption{Height of production for muon neutrinos above 1 GeV from $\pi/K$
 decay as a function of angle with different symbols. Lines show the best
 fits of Eq.~\ref{nupi} with the parameters given in Table~\ref{tab2}.
\label{fig3}
}
\end{figure}
\begin{figure}
\caption{Height of production for muon neutrinos above 1 GeV from muon
 decay as a function of angle (dots). Lines show the best fits of
 Eq.~\ref{numuapprox} with the parameters given in Table~\ref{tab2}.
\label{fig4}
}
\end{figure}
\begin{figure}
\caption{Height of production for muon neutrinos above 1, 10, and 100 GeV
 from muon decay at $\cos \theta$ = 0.25. Lines show the best fits 
of Eq.~\ref{numuapprox} and with normalization factors of 0.83, 0.50, and
 0.33 for 1, 10, and 100 GeV respectively.
\label{fig5}
}
\end{figure}

\begin{references}
%
\bibitem {SuperK} Z. Conner, Hightlight talk at the 25th International
 Cosmic Ray Conf. (Durban, South Africa, 1997) to be published;
 J.G.~Learned (SuperKamiokande Collaboration), {\it Proc. 25th
 International Cosmic Ray Conference},
 eds. M.S.~Potgieter, B.C.~Raubenheimer and D.J.~van der Walt,
 {\bf 7}, 73 (1997).
%
\bibitem {IMB}  R. Becker-Szendy   {\it et al.},  (IMB Collaboration),
{\it Phys. Rev.} {\bf D46} (1992) 3720. See also D. Casper {\it et al.}
{\it Phys. Rev. Letters}, {\bf 66} (1991) 2561.
%
\bibitem {Kam} K.S. Hirata  {\it et al.}, (Kam-II Collaboration),
{\it Phys. Letters}, {\bf B280} (1992) 146 and Y. Fukuda {\it et al.},
{\it Phys. Letters},{\bf B335} (1994) 237.  
%
\bibitem {Soud} M.C. Goodman {\it et al.}, (Soudan 2 Collaboration),
 {\it Proc. 25th International Cosmic Ray Conference},
 eds. M.S.~Potgieter, B.C.~Raubenheimer and D.J.~van der Walt,
 {\bf 7}, 77 (1997). 
%
\bibitem {BoehmV} {\em Physics of Massive Neutrinos}, by Felix Boehm \&
Peter Vogel (Cambridge University Press, 1992).
%
\bibitem {Foglietal} G.L. Fogli and E. Lisi, {\it Phys. Rev.}
, D{\bf 52} (1995) 2775.
%
\bibitem{Parkeetal} R.H.~Bernstein \& S. Parke, {\it Phys. Rev. D}, {\bf 44},
2059 (1991).
%
\bibitem {Agrawal} V. Agrawal {\it et al.}, {\it Phys. Rev.} 
D{\bf 53} (1996), 1314.
%
\bibitem {GHKLMNS} T.K. Gaisser {\it et al.}, {\it Phys. Rev.} D{\bf 54}
(1996) 5578.

%
\bibitem {BGS} Giles Barr, T.K. Gaisser \& Todor Stanev,
Phys. Rev. D39 (1989) 3532.
%
\bibitem{Lee} H. Lee \& S.A. Bludman, {\it Phys. Rev.} D{\bf 37}
(1988) 122.
%
\bibitem{conference} T.K. Gaisser \& Todor Stanev, Proc.
24th ICRC (Rome) vol. 1, p. 694.
%
\bibitem {Webber} W.R. Webber, R.L. Golden and S.A. Stephens, in {\it Proc.
20th Int. Cosm. Ray Conf.} (Moscow) {\bf 1}, 325 (1987).
%
\bibitem {Seo} E.-S. Seo {\em et al. Ap. J.} {\bf 378}, 763 (1991).
%
\bibitem {EGLS} J.~Engel {\it et al.  Phys. Rev.} {\bf D46}, 5013 (1992).
%
\bibitem{ThunIng} M. Thunmann, G. Ingelman \& P. Gondolo,
{\it Astropart. Physics} {\bf 5} (1996) 309.
%
\bibitem{Volkova} L.V. Volkova, W. Fulgione, P. Galeotti \& O. Saavedra, 
{\it Nuovo Cimento} C{\bf 10} (1987) 465.
%
\bibitem{MASS} R. Bellotti {\it et al.}, {\it Phys. Rev.} D{\bf 53},
35 (1996).
%
\bibitem {Circella} Marco Circella, Carlo De Marzo, T.K. Gaisser \&
Todor Stanev, Proc. 25th Int. Cosmic Ray Conf. (Durban) {\bf 7}, 117 (1997).
%
\bibitem {book} {\em Cosmic Rays and Particle Physics}, by Thomas K. Gaisser,
(Cambridge University Press, 1990).
%
\bibitem {Lipari} Paolo Lipari, {\em Astroparticle Physics}
{\bf 1}, 195 (1993).
%
\bibitem{SNO} M.E.~Moorhead (for the SNO Collaroration), {\it Nucl. Phys. B}
 (Proc. Suppl.) {\bf 48} 378 (1995).
\end{references}
\end{document}